\documentclass[fleqn,twoside]{article}
\usepackage{espcrc2}

\usepackage{graphicx}
\usepackage{epsfig}
\usepackage[figuresright]{rotating}

\newcommand{\defdecay}[2] {\def#1{\ifmmode{#2}\else{${#2}$}\fi}}
\defdecay{\twopic} {\pi^+ \pi^-}
\defdecay{\pipic} {\twopic}
\defdecay{\piopio} {\pi^0 \pi^0}
\defdecay{\pipin} {\piopio}

\title{The final measurement of $\epsilon'/\epsilon$ by NA48}

\author{G.~Unal\address{Laboratoire de l'Acc\'el\'erateur Lin\'eaire \\
        IN2P3-CNRS et Universit\'e Paris-Sud, \\
        B.P.34, 91898 Orsay Cedex, France}
        \thanks{\fussy on behalf of the NA48 Collaboration
  (Cagliari, Cambridge, CERN, Dubna, Edinburgh, Ferrara, Firenze,
   Mainz, Orsay, Perugia, Pisa, Saclay, Siegen, Torino, Warsaw, Wien)}}
       
\begin{document}

\begin{abstract}
The direct CP violation parameter Re($\epsilon'$/$\epsilon$) has
been measured from the decay rates of neutral kaons into two
pions using the NA48 detector at the CERN SPS. The 2001 running period
was devoted to collecting additional data under varied conditions compared
to earlier years (1997-99). The 2001 data yield the result:
Re($\epsilon$/$\epsilon'$)~=~$(13.7\pm3.1)\times10^{-4}$. Combining
this result with that published from the 1997,98 and 99 data, an overall
value of Re($\epsilon'$/$\epsilon$)~=~$(14.7\pm2.2)\times10^{-4}$ is
obtained from the NA48 experiment.
\vspace{1pc}
\end{abstract}

\maketitle

\section{INTRODUCTION}

\subsection{CP violation in the neutral kaon system}

CP violation has been discovered in the neutral kaon
system in 1964 \cite{turlay}. The main component of the
effect \cite{wolfenstein} occurs in the mixing between
$K_0$ and $\overline{K_0}$. The physical
states $K_S$ and $K_L$ deviate from pure CP = $\pm$ 1
eigenstates, with the mixing described by the
parameter $\epsilon$. Direct CP violation
can occur in kaon decays to two pions through
the interference of amplitudes with different isospins \cite{directcp}.
This is described by the
parameter $\epsilon'$. The quantity which can be measured
experimentally is the double ratio R of the decay widths: \\
\begin{eqnarray}
R & = & \frac{ \Gamma( K_L \rightarrow \pi^0 \pi^0) /  \Gamma( K_S \rightarrow \pi^0 \pi^0)}{ \Gamma( K_L \rightarrow \pi^+ \pi^-) / \Gamma( K_S \rightarrow \pi^+\pi^-)}  \nonumber \\
 & \approx & 1 - 6 \times Re(\epsilon'/\epsilon)  \nonumber
\end{eqnarray}

 In the Standard Model, CP violation arises 
from the irreducible complex
phase in the CKM matrix \cite{ckm}. Direct CP violation
is predicted by the Standard Model, with typical computations
for Re($\epsilon'/\epsilon$)
ranging from $\approx$ -10 to 30 $\times$10$^{-4}$ \cite{theory}.

From the data taken in 1997-98-99,
NA48 published~\cite{na48_99}
a result (15.3$\pm$2.6)$\times$10$^{-4}$.
KTeV~\cite{ktev} published a result based on the data taken in
1996-1997 (20.7$\pm$2.8)$\times$10$^{-4}$.
These recent results showed the existence of direct CP
violation in the neutral kaon system.

 We report here on the measurement of Re($\epsilon'/\epsilon$) performed
using the 2001 data sample, recorded in somewhat different experimental
conditions by the NA48 experiment. 

 After the 1999 data-taking period, the drift chambers of the experiment
were damaged by the implosion of the beam tube. The data
taking in 2001 took place with rebuilt drift chambers. Thanks
to the possibility of a better SPS duty cycle,  the
data could be taken at a 30\% lower beam intensity,
allowing the insensitivity of the result to intensity-related
effects to be checked,
and the statistics for the final
$\epsilon'/\epsilon$ measurement by NA48 to be completed.
The statistics accumulated during
the 2001 data-taking period is roughly half of the
total statistics accumulated in
the 1998 and 99 periods. The details of the analysis of this
data set can be found in \cite{na48_2001}.

\subsection{NA48 Method}

 The measurement of $R$ proceeds by counting the
number of events in each of the four decay modes.
The experiment is designed to exploit cancellations of systematic
effects contributing symmetrically to different components
of the R.
 Data are collected simultaneously in the four decay modes,
cancelling the absolute fluxes and minimising the sensitivity of the
measurement to accidental activity and variations of the detection
efficiency.
 The $K_L$ and $K_S$ decays are provided by two nearly collinear
beams.
To minimise the difference in acceptance due to the
large difference in average decay lengths, only $K_L$ decays occurring
in the region also populated by $K_S$ are used. The $K_L$ events 
are furthermore weighted as a function of the
proper lifetime in order to equalise the decay vertex distribution
of $K_L$ and $K_S$ events. With this procedure, the acceptance
correction cancels at first order in R, thus
minimising the systematic uncertainty from Monte Carlo modelling of
the experiment.

\section{BEAMS AND DETECTOR}

\subsection{Beams}

 The neutral beams are derived from 400 GeV/$c$ protons extracted
from the CERN SPS. Because of the different mean decay lengths
of $K_L$ and $K_S$,
two different production targets
are used, located 126~m and 6~m upstream of the beginning of the decay
region. For each SPS pulse (5.2~s every 16.8~s), $\approx$2.4$\times$10$^{12}$
protons hit the $K_L$ production target. Three stages of collimation are
used to define the $K_L$ beam, at a production angle of 2.4~mrad.
 Part of the non-interacting protons impinge on a bent silicon mono-crystal.
A small fraction undergoes channelling and produces a proton beam
of $\approx$5$\times$10$^{7}$ protons per pulse transported to the
$K_S$ production target, where a second neutral beam is derived, at
a production angle of 4.2~mrad. The $K_S$ beam enters the fiducial
decay region 6.8~cm above the $K_L$ beam. The beams converge
with an angle of 0.6~mrad and the axes of the two beams cross at
the position of the electromagnetic calorimeter.
 To distinguish $K_L$ and $K_S$ decays, the protons directed
to the $K_S$ target are detected by an array of scintillation counters
which comprise the tagging detector. The time of each proton is recorded and
is compared with the time of the decay as measured in the main detector.
The presence (absence) of a proton in coincidence with the event defines
the event as $K_S$ ($K_L$).
 The $K_S$ beam traverses an anti-counter (AKS), formed by a set
of scintillation counters following a 3~mm thick iridium crystal.
This detector provides an accurate definition of the beginning
of the decay region where it is located by vetoing $K_S$ decays
occurring upstream.

\subsection{Main detector}

Charged pion decays are measured by a magnetic spectrometer
comprised of four drift chambers and a dipole magnet giving a momentum
kick of 265 MeV/$c$.
The momentum resolution is $\sigma_p/p$ = $0.48\% \oplus 0.009\% \times p$
($p$ in GeV/$c$). Two plastic scintillator hodoscope planes are
located after the spectrometer. They are used to determine the
event time of the charged events for the tagging procedure.

 A quasi-homogeneous liquid krypton electromagnetic calorimeter
with a projective tower readout is used to measure the photons
from $\pi^0 \pi^0$ events. 
The
13212 readout cells have each a cross-section of $\approx$ 2$\times$2 cm$^2$.
The energy resolution is
$\sigma(E)/E$ = $0.032/\sqrt{E} \oplus 0.09/E \oplus 0.0042$, where $E$
is in GeV. The spatial resolution is better than 1~mm above 25 GeV.
This detector is also used to measure the time of the photons for
the tagging procedure.

 Located after the electromagnetic calorimeter are a iron-scintillator
hadron calorimeter, followed by a muon counter consisting of
three planes of scintillators sandwiched between 80~cm thick iron walls.

\subsection{Trigger}

 The electromagnetic calorimeter is used to trigger on $\pi^0\pi^0$
events. At the trigger level, the calorimeter data are reduced
to $x$ and $y$ projections which are used to reconstruct the
total energy as well as to estimate the decay position of the
event. The efficiency of the $\pi^0\pi^0$ trigger is
$(99.901\pm0.015)\%$ and is $K_S$-$K_L$ symmetric.
 A two level trigger is used for $\pi^+\pi^-$ decays. At first
level, the hodoscope is placed in coincidence with a total energy
condition defined using both calorimeters. The second level trigger
uses information from the drift chambers to perform a fast
event reconstruction. The efficiency of the  $\pi^+\pi^-$ trigger
is $(98.697\pm0.017)\%$. This is higher by $0.9\%$ than for
the 1998-99 data thanks mostly to the lower beam intensity. The
efficiency is measured separately for $K_S$ and $K_L$ and the
double ratio is corrected for the small difference.

\section{EVENT RECONSTRUCTION AND SELECTION}

 $K \rightarrow \pi^0\pi^0$ decays are selected using only
data from the LKr calorimeter. The details of the reconstruction
and of the cuts applied can be found
in \cite{na48_2001} and \cite{na48_99}. From the measured
energies and impact point positions on the calorimeter of the
four showers, the decay vertex position along the beam axis is
computed assuming that their invariant mass is
the kaon mass. The invariant mass of the two photons pairs are then
computed (the resolution is better than 1~MeV/$c^2$)
and compared to the nominal $\pi^0$ mass, constructing
a $\chi^2$ variable. To reject
the residual background from $K_L \rightarrow 3\pi^0$ events, a
cut on $\chi^2$ is applied. 
This residual background fraction is
$(5.6\pm2.0)\times10^{-4}$ in the $K_L$ sample, while the $K_S$ sample
is background free. The background from $K_S$ decays produced
by scattering of beam particles in the collimators in the $K_L$ beam
is $(8.8\pm2.0)\times10^{-4}$.

 $K \rightarrow \pi^+\pi^-$ decays are reconstructed from tracks in the
spectrometer. In the $\pi^+\pi^-$ mode, both the longitudinal
and the transverse decay vertex position can be reconstructed, allowing
a clean identification of $K_S$ and $K_L$ decays.
 A cut is applied on the ratio of the two track momenta to remove
asymmetric decays in which one of the tracks could be close to the
beam tube where the Monte Carlo modelling is more critical.
 To reject background from semileptonic $K_L$ decays, events with
tracks consistent with being either an electron 
(from the E/p ratio of the energy deposited in the LKr calorimeter over
the track momenta) or a muon (using hits in the muon counters) are rejected.
 Kinematical cuts are also applied on the $\pi^+\pi^-$ invariant mass
(the resolution is typically 2.5~MeV/$c^2$) and on the reconstructed transverse
momentum. The residual background in the $K_L$ sample is $(14.2\pm3.0)\times10^{-4}$,
while it is negligible in the $K_S$ sample.

 The fiducial ranges in kaon energy $E_K$ and in proper time $\tau$ used
to count events are chosen to be $70< E_K < 170$~GeV and
$0 < \tau < 3.5\ \tau_S$, where $\tau=0$ is defined at the position
of the AKS counter and $\tau_S$ is the $K_S$ mean lifetime.
For $K_L$ events, the decay time cut is applied on reconstructed $\tau$,
while for $K_S$ events the cut at $\tau=0$ is applied
using the AKS to veto decays occurring upstream.
 The determinations of the kaon energy, the decay vertex and the proper
time in the \pipin\ mode rely on measurements of the photon energies and
positions with the calorimeter. The uniformity of the calorimeter
response is optimised using $K_{e3}$ decays and checked using $\pi^0$
and $\eta$ decays produced during special (``$\eta$'') runs in which a $\pi^-$
beam strikes two thin targets located near the beginning and the end
of the fiducial decay region. The absolute energy scale is adjusted,
with a 0.03\% accuracy,
using $K_S\rightarrow\pi^0\pi^0$ decays, such that the reconstructed
AKS position matches the true value. It is also checked using data from the
$\eta$ runs (the $\eta$ mass is taken from \cite{eta_mass}). 
Taking all uncertainties into account (including non-linearities
in the energy response), the total systematic uncertainty on R
from the reconstruction of \pipin\ events is found to be $\pm5.3\times10^{-4}$
\cite{na48_2001}.
 The uncertainty on R from the reconstruction of
\pipic\ events is $\pm2.8\times10^{-4}$.

\begin{figure}[htb]
\mbox{\epsfig{file=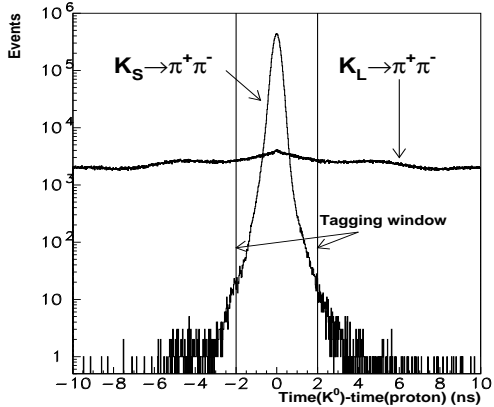,height=6cm,width=7cm}}
\caption{Time coincidence for $K_S$ and $K_L$ \pipic decays, identified
by their reconstructed vertex position.}
\label{fig:tagging}
\end{figure}

\begin{figure}[htb]
\mbox{\epsfig{file=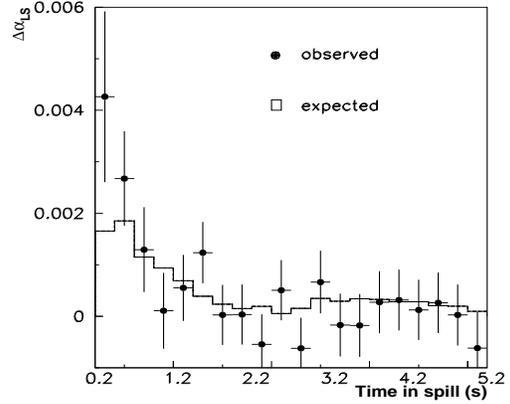,height=6cm,width=7cm}}
\caption{Measured compared to predicted values of $\Delta \alpha_{LS}$
as a function of the time during the spill.}
\label{fig:dals_expected}
\end{figure}

 A decay is labelled $K_S$ is a coincidence is found within a
$\pm$~2~ns interval between its event time and a proton time
measured with the tagger. Figure \ref{fig:tagging} shows the time
distribution for $K_S$ and $K_L$ decays to \pipic\ which have been
identified by their vertex positions. From this, the misidentification
probabilities in the \pipic\ mode can be deduced:
$\alpha_{SL} = (1.12\pm0.03)\times10^{-4}$ for the $K_S$ tagging
inefficiency and $\alpha_{LS} = (8.115\pm0.010)\%$ for the $K_L$ mistagging
as $K_S$ due to an accidental coincidence between the event and a proton.
 As the same tagging procedure is used for \pipin\ and \pipic\ events,
R is only sensitive to differences in misidentification
probabilities between the \pipin\ and \pipic\ modes. Several methods
have been developed to derive these differences directly from the
data \cite{na48_2001}. The results are 
$\Delta \alpha_{SL} = (0\pm0.5)\times10^{-4}$ and
$\Delta \alpha_{LS} = (3.4\pm1.4)\times10^{-4}$ corresponding respectively
to an uncertainty on R of $\pm3.0\times10^{-4}$ and a correction
of $(6.9\pm2.8)\times10^{-4}$. The origin of $\Delta \alpha_{LS}$ is related
to a higher loss of \pipic\ events from accidental activity in the
beam and can be predicted using a technique of overlaying ``random''
events (taken proportionally to the beam intensity) with $\pi\pi$
decays to estimate these losses. The agreement between the observed
and the predicted $\Delta \alpha_{LS}$ values is illustrated in
Figure~\ref{fig:dals_expected}.

\section{ACCIDENTAL EFFECTS}

\begin{figure}[htb]
\mbox{\epsfig{file=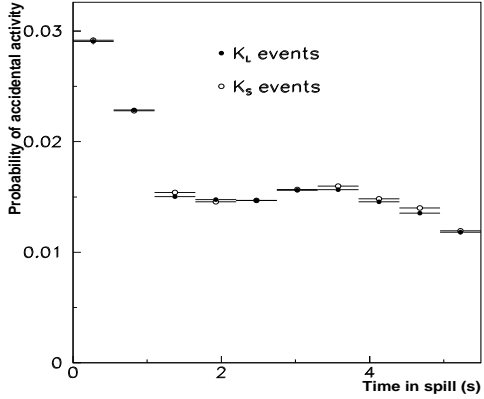,height=6cm,width=7cm}}
\caption{Probability of accidental activity in the LKr in the $\approx$~150~ns
readout window as a function of the time during the spill,
for $K_S$ and $K_L$ decays to \pipic.}
\label{fig:acci_spill2001}
\end{figure}

 The 2001 data were taken at lower beam intensity to reduce
the uncertainties related to accidental effects.
 The overlap of extra particles related to kaon decays in the
high intensity $K_L$ beam with a good event may result in the loss
of the event. The effect on R is minimised by the simultaneous
data collection in the four channels. The possible residual
effect on R can be separated into two components:

 1) A difference between the beam intensities seen
by $K_S$ and $K_L$ events: $\Delta R = \Delta \lambda
\times \Delta I/I$, where $\Delta \lambda$ is the difference
between the event losses in the \pipic\ and \pipin\ modes and
$\Delta I/I$ the difference between the $K_L$ beam intensity
seen by $K_L$ and $K_S$ events. $\Delta \lambda$ is minimised
by applying to all events the recorded dead time conditions.
The largest one is an ``overflow'' condition in the drift
chamber readout, which rejects 11\% of the $\pi\pi$ events
(this is significantly smaller than the 20\% loss in the 1998-99 sample
thanks to the lower beam instantaneous intensity and a lower
noise in the drift chambers). $\Delta \lambda$ is estimated mainly
from the overlay technique and is found to be $(1.0\pm0.5)\%$.
$\Delta I/I$ is measured from the rate of out-of-time LKr
clusters and tracks in good £ $K_S$ and $K_L$ decays. This
is illustrated in Figure~\ref{fig:acci_spill2001}. Beam monitors,
which have been improved for the 2001 data-taking period, are
also used as cross-check.
The result is $\Delta I/I = (0\pm1)\%$. The uncertainty on
R from this effect is thus $\pm1.1\times10^{-4}$. This is 
significantly better than for the 1998-99 sample, thanks to the
lower intensity and to better beam intensity monitors.

 2) A difference in illumination between $K_S$ and $K_L$
decays coupled to a variation of the event loss with the
impact points of the $K^0$ decay products. This effect is
also computed from the overlay samples. No effect is found
and the uncertainty on R is $\pm3.0\times10^{-4}$, from
the statistics of the overlay samples.

\section{RESULT AND CONCLUSIONS}

 Table~\ref{tab:result} shows the summary of the corrections
and systematic uncertainties on R for 2001 data. The residual acceptance
correction is related to the small angle between the
$K_L$ and $K_S$ beams and is computed using a large statistics
Monte Carlo sample. As it is mostly given by the beam geometry,
it does not rely on a detailed simulation of the detector.
As expected from the design of the experiment, all corrections
are small. Some systematic uncertainties are directly given by
the statistics of the control samples used to study them.
 From the 2001 sample (comprising $1.55\times10^6$
$K_L \rightarrow \pi^0\pi^0$ events, $2.16\times10^6$
$K_S \rightarrow \pi^0\pi^0$, $7.14\times10^6$
$K_L \rightarrow \pi^+\pi^-$ and $9.61\times10^6$
$K_S \rightarrow \pi^+\pi^-$), the result
R~=~$0.99181\pm0.00147\pm0.00110$ is obtained (where the
first error is statistical and the second systematic).
The corresponding value of Re($\epsilon'/\epsilon$) is
Re($\epsilon'/\epsilon$)~=~$(13.7\pm2.5\pm1.8)\times10^{-4}$.
The agreement between this new and the earlier results
is particularly significant
since they were obtained from data taken at different average beam
intensities.
Taking into account the correlated systematic
uncertainty of $\pm1.4\times10^{-4}$, the final combined result from
the NA48 experiment is 
Re($\epsilon'/\epsilon$)~=~$(14.7\pm2.2)\times10^{-4}$.

\hspace{-1cm}
\begin{table}[!htb]
\begin{center}
\caption{Corrections and systematic uncertainties on R}
\label{tab:result}
\begin{tabular}{lrrr} \hline
&\multicolumn{3}{c}{in 10$^{-4}$} \\ \hline
\pipic\ trigger                                 &  $+5.2$       & $\pm$3.6  & (stat) \\
AKS inefficiency                                &  $+1.2$       & $\pm$0.3  & \\
Reconstruction \begin{tabular}{@{}l} of \pipin\ \\ of \pipic\ \end{tabular} &
\begin{tabular}{r@{}}    ---   \\  ---   \end{tabular} &
\begin{tabular}{r@{}}   $\pm$ 5.3 \\  $\pm$ 2.8  \end{tabular} & \\
Background \begin{tabular}{@{}l} to \pipin\ \\ to \pipic\ \end{tabular} &
\begin{tabular}{r@{}} $-5.6$ \\ $+14.2$ \end{tabular} &
\begin{tabular}{r@{}} $\pm$ 2.0  \\ $\pm$ 3.0  \end{tabular} & \\
Beam scattering                                 &  $-8.8$       & $\pm$2.0  & \\
Accidental tagging                              &  $+6.9$       & $\pm$2.8  & (stat)  \\
Tagging inefficiency                            &  ---          & $\pm$3.0  & \\
Acceptance \begin{tabular}{@{}l} statistical \\ systematic \end{tabular}
                              & $+21.9$ &
     \begin{tabular}{r@{}} $\pm$ 3.5 \\ $\pm$ 4.0 \end{tabular}  & 
 \begin{tabular}{r@{}} (stat) \\ \ \end{tabular} \\
Accidental effect \begin{tabular}{@{}l} intensity  \\
                                          illumination \end{tabular}
                            &  ---    &   \begin{tabular}{r@{}} $\pm$1.1 \\
                                          $\pm$3.0 \end{tabular}  &
\begin{tabular}{r@{}} \\  (stat) \end{tabular} \\
$K_S$ in time activity      & ---     & $\pm$1.0 & \\
\hline
Total                                           & $+35.0$       & $\pm$11.0 & \\
\hline
\end{tabular}
\end{center}
\end{table}

\end{document}